\documentclass[aps,twocolumn,showpacs]{revtex4}
\usepackage{graphicx}

\begin{document}

\title{Critical Temperature for $\alpha$-Particle Condensation within a 
Momentum Projected Mean Field Approach}

\author{T. Sogo, G. R\"opke}

\affiliation{
Institut f\"ur Physik, Universit\"at Rostock, D-18051 Rostock, Germany}

\author{P. Schuck}
\affiliation{
Institut de Physique Nucl\'eaire, CNRS, UMR 8608, Orsay F-91406, France}
\affiliation{Universit\'e Paris-Sud, Orsay F-91505, France}
\affiliation{
Laboratoire de Physique et Mod\'elisation 
des Milieux Condens\'es, CNRS and Universit\'e Joseph Fourier, 
25 Avenue des Martyrs, Bo\^ite Postale 166, F-38042 Grenoble Cedex 9, France}

\begin{abstract}
Alpha-particle (quartet) condensation in homogeneous spin-isospin symmetric 
nuclear matter is investigated. 
The usual Thouless criterion for the critical temperature is 
extended to the quartet case. 
The in-medium four-body problem is strongly simplified 
by the use of a momentum projected mean field ansatz for the quartet. 
The self-consistent single particle wave functions 
are shown and discussed for various values of the density 
at the critical temperature. 
\end{abstract}

\pacs{03.75.Nt, 03.75.Ss, 21.65.-f, 74.20.Fg}

\maketitle

\section{Introduction}

Investigation of pairing in different Fermi systems 
is still on the forefront of active research. 
Examples are nuclear physics 
\cite{sb08} 
and the physics of cold fermionic atoms 
\cite{gps08}. 
The formation and condensation of heavier clusters 
in Fermi systems is much less studied.

In cold atom physics the recent advent of trapping 
three different species of fermions \cite{olk08} 
has opened up the possibility of creating gases of heavier clusters. 
For the time being those may be trions 
(bound state of three different fermions) but in 
the future one also can think of quartets (bound state of four different 
fermions). The latter are specially interesting because of their bosonic 
nature and the possibility of Bose-Einstein condensation 
(BEC) of quartets. 
The description of quartet condensation to occur 
has been attempted with an extension of the so-called Cooper problem 
to the four body case in \cite{km05}.
In \cite{sg99} a variational procedure for the condensation 
of ($2s+1$)-component fermion clusters, 
with $s$ the fermion spin, has been proposed.
A quartet phase has been found in a one dimensional model 
with four different fermions \cite{rcl08}.

On the other hand,  nuclear physics, because it is a four-component 
fermion-system (proton/neutron spin up/down), all fermions 
attracting one another, 
leading to the very strongly bound $\alpha$-particle, is a proto-type 
system for quartetting. There, the formation 
of clusters has been an object of study almost since the beginning of nuclear 
physics \cite{www37}. Of course, pairing also exists in nuclei from 
where it is concluded that neutron stars are superfluid. 
Nuclei are very small quantum 
objects with only a (slowly) fluctuating phase (the conjugate variable to 
particle number $N$). Actually the number of Cooper 
pairs in nuclei generally does not exceed  about a dozen (often much less) and 
yet clear signs of superfluidity are observed in nuclei (e.g., moments of 
inertia strongly reduced from their classical value), implying that no 
critical size exists from where signatures of superfluidity abruptly 
disappear. 
One, thus, can safely extrapolate from finite nuclei to superfluidity in 
neutron stars. On the other hand, as already mentioned, in nuclear physics 
the existence of quartets ($\alpha$-particles) as subclusters of nuclei is 
omnipresent. As well known, many lighter nuclei with equal 
proton and neutron numbers ($Z=N$) show, 
for instance in excited states, strong $\alpha$ clustering. 
The concept that these $\alpha$-particles may form a condensate 
in certain low density states of nuclei and that this may, 
in analogy to the pairing case, 
be a precursor sign of $\alpha$-particle condensation in infinite matter
\cite{rss98}, 
has come up only recently \cite{thsr01}. 
Also heavy 
nuclei seem to have preformed $\alpha$-clusters in the surface because of 
their well known spontaneous $\alpha$ decay properties.

Symmetric nuclear matter does not exist in nature because of the too strong 
Coulomb repulsion. However, in collapsing stars, so-called proto-neutron 
stars, the fraction of protons is still high \cite{st83} and the formation of 
$\alpha$-particles and, at sufficiently low temperature, their condensation 
may eventually be possible. At any rate, it seems evident that nuclear 
matter at various 
degrees of asymmetry is unstable with respect to cluster formation in the 
low density regime. At zero temperature, the most stable nucleus is $^{56}$Fe 
but as a function of temperature, density, asymmetry, other cluster 
compositions of infinite baryonic matter may be formed. Several theoretical 
studies predict $\alpha$-phases to exist in certain temperature-density-
asymmetry domains \cite{sto98}.

In view of the complexity of the task, the objective of the present work is 
quite modest. We want to study the critical temperature of $\alpha$-particle 
condensation as a function of density and temperature in symmetric nuclear 
matter. Still, even this task will not be carried out down to the BEC limit. 
We will study the critical temperature $T_c^{\alpha}$ for the onset of 
formation of $\alpha$-particles in a thermal gas of nucleons. This shall be 
done with a theory analogous to the famous Thouless criterion for the onset 
of formation of Cooper pairs in a superconductor. On the microscopic level the 
problem is still very challenging, since it amounts to solve an in-medium 
four-body problem. 
In spite of that, solutions have already been worked out in the 
past, either solving the Faddeev-Yakubovsky equations \cite{beyer} or with an 
approximate procedure \cite{rss98}.

In this work, we will continue along those lines. The final objective is to 
reach the BEC regime in a treatment similar to the one of 
Nozi\`eres and Schmitt-Rink (NSR) theory \cite{nsr85}, but for quartets.
Needless to 
say that this only will be possible if the whole formalism can radically be 
simplified. Actually, as we will show in this work, such a procedure may well 
exist. In any case, it is not conceivable that one treats condensation of 
bosonic clusters built out of $N$ fermions on the level of non-linear 
in medium $N$-body 
equations for $N>2$. On the other hand, it is well known, 
that nuclei can satisfactorily be described, in mean field 
approximation \cite{rs00}. 
Projecting these mean field (Hartree-Fock) type of solutions on zero 
total momentum (${\bf K}=0$) will then allow these mean field clusters to Bose 
condense. Actually it is well known among the nuclear physics community that 
even for such a small nucleus as the $\alpha$-particle a momentum projected 
mean field approach yields a very reasonable description \cite{berger}. 
The reason for this stems, as already mentioned, from the presence of 
four different fermions, all attracting one another with about the same force.

\begin{figure}
\includegraphics[width=40mm]{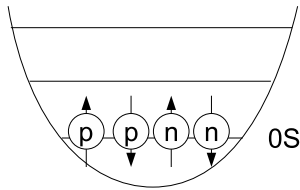}
\caption{\label{fig1}
Sketch of $\alpha$-particle configuration, 
indicating that the two protons and two neutrons 
occupy the lowest $0S$ level 
in the mean field potential of harmonic oscillator shape.}
\end{figure}

In Fig.~\ref{fig1} 
we sketch the situation, indicating that the two protons and two neutrons 
occupy the lowest $0S$ level of the mean field potential. Actually 
calculations show that the $0S$ orbital of the self-consistent mean field 
resembles very much an oscillator wave function of Gaussian shape. In this 
respect the cartoon in Fig.~\ref{fig1} 
is not so far from reality. We suspect that 
the situation is generic for all strongly bound quartets which may be 
produced in the future and, therefore, our present study is of quite general 
interest. We will adopt this momentum projected mean field 
procedure in this work.

\section{The in-medium four body equation}

In-medium four body equations are well documented in the literature since long 
\cite{ds94}. In the present case of an in-medium 
quartet, the corresponding equation reads as follows \cite{rss98}:
\begin{eqnarray}
(E-\varepsilon_{1234})\Psi_{1234}
&=&(1-f_1-f_2)\sum_{1'2'}v_{12,1'2'}\Psi_{1'2'34}\nonumber \\
&+& (1-f_1-f_3)\sum_{1'3'}v_{13,1'3'}\Psi_{1'23'4}\nonumber \\
&+&{\rm permutations},
\label{eq1}
\end{eqnarray}
where $f_i=f(\varepsilon_i)=[e^{(\varepsilon_i-\mu)/T}+1]^{-1}$ 
with $\varepsilon_i=\varepsilon(k_i)=k_i^2/(2m)$ 
is the Fermi-Dirac distribution and 
$\varepsilon_{1234}=\varepsilon_{1}+\varepsilon_{2}
+\varepsilon_{3}+\varepsilon_{4}$ 
($\hbar=c=k_B=1$: natural units). 
The matrix element of the interaction is 
$v_{12, 1'2'}$ with the numbers 1, 2, 3, $\cdots$ 
standing for all quantum numbers as momenta, spin, isospin, etc.,
as also in all other quantities in (\ref{eq1}).

In Eq.~(\ref{eq1}), when $E=4\mu$, this signals quartet 
condensation in very much the same manner as in the two body equation
\begin{equation}
(E - \varepsilon_{12})\Psi_{12} = ( 1- f_1 -f_2)
\sum_{1'2'}v_{12,1'2'}\Psi_{1'2'},
\label{eq2}
\end{equation}
where $\varepsilon_{12}=\varepsilon_{1}+\varepsilon_{2}$,
the approach of $T \rightarrow T_c$ such that $E \rightarrow 2\mu$ signals the 
transition to a superconducting or superfluid state (the well known Thouless 
criterion \cite{thouless60}).

Of course, as already stated several times, the determination 
of $T_c^{\alpha}$  needs the heavy solution of the in-medium modified 
four particle equation (\ref{eq1}).

Following the discussion in the introduction, we, therefore, make the 
following 'projected' mean field ansatz for the quartet wave function 
\cite{km05,schuck08,sg99}, 
\begin{equation}
\Psi_{1234}= (2\pi)^3
\delta^{(3)}({\bf k}_1 +{\bf k}_2 + {\bf k}_3 + {\bf k}_4) 
\prod_{i=1}^4\varphi({\bf k}_i)\chi^{ST},
\label{eq3}
\end{equation}

\noindent
where $\chi^{ST}$ is the spin-isospin function which we suppose to be the 
one of a scalar ($S=T=0$). 
We will not further mention it from now on. 
We work in momentum space and 
$\varphi({\bf k})$ is the as-yet unknown single particle $0S$ wave function. 
In position space, this leads to the usual formula 
\cite{rs00} $\Psi_{1234} \rightarrow \int d^3R \prod_{i=1}^4 
\tilde\varphi({\bf r}_i - {\bf R})$ 
where $\tilde\varphi({\bf r}_i)$ is the Fourier 
transform of $\varphi({\bf k}_i)$. If we take for $\varphi({\bf k}_i)$ 
Gaussian shape, this gives: 
$\Psi_{1234} \rightarrow \exp[-c\sum_{1\leq i<k \leq 4} 
({\bf r}_i - {\bf r}_k)^2]$
 which is the translationally invariant ansatz often used to describe 
$\alpha$-clusters in nuclei. For instance, it is also employed 
in the $\alpha$-particle condensate wave function 
of Tohsaki, Horiuchi, Schuck, R\"opke (THSR) in \cite{thsr01}.

Inserting the ansatz (\ref{eq3}) into (\ref{eq1}) and integrating 
over superfluous 
variables, or minimizing the energy, we arrive at the following non-linear, 
Hartree-Fock type of 
equation for the single particle $0S$ wave function 
$\varphi(k)=\varphi(|{\bf k}|)$

\begin{equation}
A(k)\varphi(k) + 3B(k)+ 3C(k)\varphi(k)=0,
\label{eq4}
\end{equation}
where $A(k)$, $B(k)$, and $C(k)$ are given by:
\begin{eqnarray}
{\cal A}(k_1)
&=&\int \prod^{4}_{i=2}\frac{d^3 k_i}{(2\pi)^3}
\left[\sum_{i=1}^4\frac{k_i^2}{2m} - 4\mu\right] \nonumber \\
&\times&
|\varphi(k_2)|^2|\varphi(k_3)|^2|\varphi(k_4)|^2
(2\pi)^3\delta^{(3)}(\sum_{i=1}^4{\bf k}_i) 
\label{eq4c}\\
%%%
{\cal B}(k_1)&=&\int \prod^{4}_{i=2}\frac{d^3 k_i}{(2\pi)^3}
\frac{d^3k_1'}{(2\pi)^3}\frac{d^3k_2'}{(2\pi)^3}
(1-f(\varepsilon_1)-f(\varepsilon_2))
\nonumber \\
&\times& v_{{\bf k}_1 {\bf k}_2,{\bf k}_1' {\bf k}_2'}
\varphi(k_1')\varphi(k_2')\varphi(k_2)|\varphi(k_3)|^2|\varphi(k_4)|^2
\nonumber \\
&\times& (2\pi)^3\delta^{(3)}(\sum_{i=1}^4{\bf k}_i)
\label{eq4a}\\
%%%
{\cal C}(k_1)
&=&\int \prod^{4}_{i=2}\frac{d^3 k_i}{(2\pi)^3} 
\frac{d^3k_2'}{(2\pi)^3}\frac{d^3k_3'}{(2\pi)^3}
(1-f(\varepsilon_2)-f(\varepsilon_3))
\nonumber \\
&\times&v_{{\bf k}_2 {\bf k}_3, {\bf k}_2' {\bf k}_3'} 
\varphi(k_2)\varphi(k_2')\varphi(k_3)\varphi(k_3')|\varphi(k_4)|^2 
\nonumber \\
&\times& (2\pi)^3\delta^{(3)}(\sum_{i=1}^4{\bf k}_i),
\label{eq4b}
\end{eqnarray}
where, in this pilot study, we neglect mean 
field shifts and effective mass contributions.

From Eq.~(\ref{eq4}), we obtain the single particle wave function
in momentum space as
\begin{equation}
\varphi(k)=\frac{-3B(k)}{A(k)+3C(k)}.
\label{eq9}
\end{equation}
As seen in Eqs.~(\ref{eq4c}), (\ref{eq4a}), and (\ref{eq4b}), 
since $A(k)$, $B(k)$, and $C(k)$ depend on 
the wave function of $\varphi(k)$,
Eq.~(\ref{eq9}) is strongly non-linear. Its solution can be found by iteration.

%Still the equation to be solved is rather complicated for a general two 
%body force $v_{{\bf k}_1 {\bf k}_2, {\bf k}'_1 {\bf k}'_2}$. 
For a general two body force 
$v_{{\bf k}_1 {\bf k}_2, {\bf k}'_1 {\bf k}'_2}$,
the equation to be solved is still rather complicated.
We, therefore, proceed to the last 
simplification and replace the two body force by a unique separable one, 
that is 
\begin{equation}
v_{{\bf k}_1 {\bf k}_2, {\bf k}'_1 {\bf k}'_2} = \lambda
e^{-k^2/k_0^2}e^{-k'^2/k_0^2}
(2\pi)^3\delta^{(3)}({\bf K}-{\bf K}'),
\label{eq5}
\end{equation}
where ${\bf k}=({\bf k}_1-{\bf k}_2)/2$, 
${\bf k}'=({\bf k}_1'-{\bf k}_2')/2$, 
${\bf K}={\bf k}_1+{\bf k}_2$, and ${\bf K}'={\bf k}_1'+{\bf k}_2'$.
This means that we take a spin-isospin averaged two body interaction and 
disregard that in principle the force may be somewhat different in the  
$S,T = 0, 1$ or $1, 0$ channels.

We are now ready to study the solution of (\ref{eq1}) for the critical 
temperature $T_c^{\alpha}$ when the eigenvalue hits $4\mu$.
For later comparison, the deuteron (pair) wave function at the 
critical temperature
is also represented from Eqs.~(\ref{eq2}) and (\ref{eq5}) to be
\begin{equation}
\phi(k)= -\frac{1-2f(\varepsilon)}{k^2/m-2\mu}\lambda e^{-k^2/k_0^2}
\int \frac{d^3k'}{(2\pi)^3} e^{-k^2/k_0^2} \phi(k'),
\label{eq8}
\end{equation}
where $\phi(k)$ is a relative wave function of two particles given by
$\Psi_{12} \to \phi(|\frac{{\bf k}_1-{\bf k}_2}{2}|)
\delta^{(3)}({\bf k}_1+{\bf k}_2)$, and 
$\varepsilon=k^2/(2m)$.
From Eq.~(\ref{eq8}), the critical temperature of pair condensation
is obtained with the following equation:
\begin{equation}
1=-\lambda \int \frac{d^3k}{(2\pi)^3}
\frac{1-2f(\varepsilon)}{k^2/m-2\mu} e^{-2k^2/k_0^2}.
\label{eq10}
\end{equation}

\section{ Results for the critical temperature $T_c^{\alpha}$}

\begin{figure}
\begin{center}
\includegraphics[width=60mm]{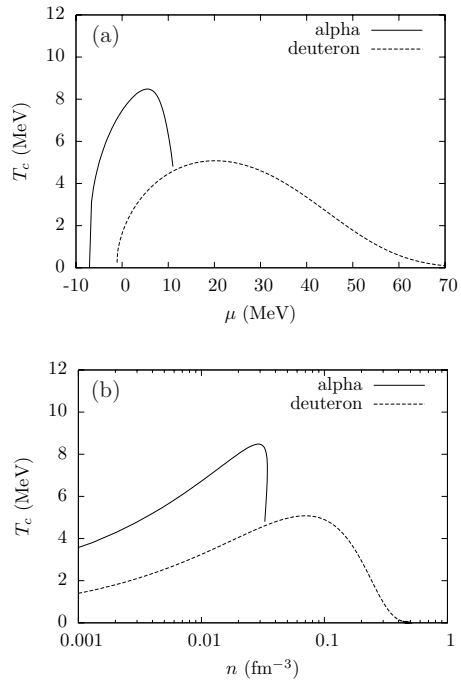}
\end{center}
\caption{\label{fig2}
Critical temperature of alpha and deuteron condensations
as functions of chemical potential (a) and density (b),
derived from Eq.~(\ref{eq4}) for the $\alpha$-particle 
and Eq.~(\ref{eq10}) for the deuteron,
respectively.}
\end{figure}

In order to determine the critical temperature for 
$\alpha$-condensation as a function of density $n$, 
we need to determine the chemical potential $\mu$ via 
%$n=4\int d^3k f(\varepsilon)/(2\pi)^3$ 
\begin{equation}
n=4\int \frac{d^3k}{(2\pi)^3} f(\varepsilon)
\label{eq-density}
\end{equation}
and adjust the temperature so that the 
eigenvalue of (\ref{eq1}) hits $4\mu$. 
The two open constants $\lambda$ and $k_0$ in Eq. (\ref{eq5}) are determined 
so that binding energy ($-28.3$ MeV) and radius 
($1.71$ fm) of the free ( $f_i=0$) $\alpha$-particle come out right.
The adjusted values are: $\lambda=-992$ MeV fm$^{3}$, and 
$b=1.43$ fm$^{-1}$.
The results of the calculation 
are shown in Fig.~\ref{fig2}.

In Fig. \ref{fig2}, the maximum of critical temperature 
$T^{\alpha}_{c, {\rm max}}$
is at $\mu=5.5$ MeV, and 
the $\alpha$-condensation can exist up to  $\mu_{\rm max}=11$ MeV.  
It is very remarkable that the obtained results for $T_c^{\alpha}$ 
well agree with a direct solution of (1) \cite{beyer}. 
These results for $T_c^{\alpha}$ are by about 25 percent higher than the ones 
of our earlier publication \cite{rss98}. We, however, checked that the 
underlying radius of the $\alpha$-particle in that work is  
larger than the experimental value and that $T_c^{\alpha}$ 
decreases with increasing radius of 
$\alpha$-particle. 
Furthermore a different variational wave function was used in 
\cite{rss98}.

In Fig.~\ref{fig2} 
we also show the critical temperature for deuteron condensation
derived from Eq.~(\ref{eq10}). 
In this case, 
we take $\lambda= -1305$ MeV fm$^3$ and $k_0 = 1.46$ fm$^{-1}$ to 
get experimental energy ($-2.2$ MeV) and radius ($1.95$ fm) of the deuteron. 
It is seen that at higher densities 
deuteron condensation wins over the one of $\alpha$-particle. 
The latter breaks down rather abruptly at a critical positive value 
of the chemical potential. Roughly speaking, this corresponds to the point 
where the 
$\alpha$-particles start to overlap. This behavior stems from the fact 
that Fermi-Dirac distributions 
in the four body case, see (\ref{eq1}), can never become step-like, as in the 
two body case, even not at zero temperature, 
since the pairs in an $\alpha$-particle 
are always in motion. As a consequence,  
$\alpha$-condensation generally only 
exists as a BEC phase and the weak coupling regime is absent.

\begin{figure}
\begin{center}
\includegraphics[width=80mm]{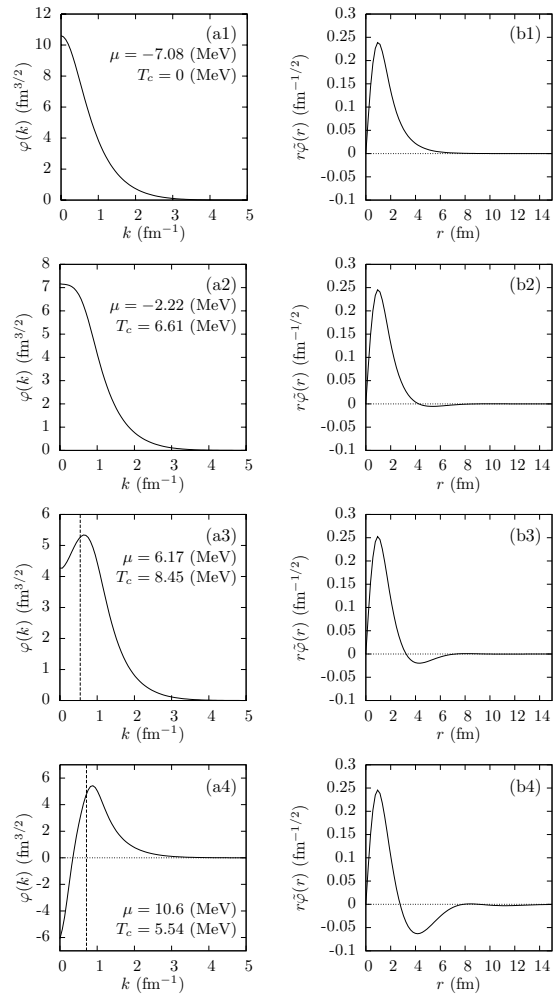}
\end{center}
\caption{\label{fig3}
Single particle wave functions 
in momentum space $\varphi(k)$ (a), 
and in position space $r\tilde \varphi(r)$ (b) at critical temperature,
Eq.~(\ref{eq9}). 
From top to bottom: 
(1) $\mu=-7.08$ MeV, $T_c=0$ MeV, $n=0$ fm$^{-3}$
(2) $\mu=-2.22$ MeV, $T_c=6.61$ MeV, $n=9.41 \times 10^{-3}$ fm$^{-3}$,
(3) $\mu=6.17$ MeV, $T_c=8.45$ MeV, $n=3.07 \times 10^{-2}$ fm$^{-3}$,
and 
(4) $\mu=10.6$ MeV, $T_c=5.54$ MeV, $n=3.34 \times 10^{-2}$ fm$^{-3}$.
Figs.~(a1) and (b1) correspond to the wave functions 
for free $\alpha$-particle.
The vertical lines in (a3) and (a4) 
are at the Fermi wave length $k_F=\sqrt{2m\mu}$.}
\end{figure}

Fig.~\ref{fig3} shows the normalized 
self-consistent solution of the wave function 
in momentum space derived 
from Eq.~(\ref{eq9}) and the wave function in position space defined by
its Fourier transform $\tilde \varphi(r)$.
Fig.~\ref{fig3}-(a1) and (b1) are the wave functions 
of the free $\alpha$-particle.
As discussed in the Introduction, 
the wave function resembles a Gaussian and
this shape is approximately maintained as long as $\mu$ is negative,
see Fig.~\ref{fig3}-(a2).
On the contrary,
the wave function of Fig.~\ref{fig3}-(a3),
where the chemical potential is positive, has a dip around $k=0$ which is
due to the Pauli blocking effect.
For the even larger positive chemical potential
of Fig.~\ref{fig3}-(a4) the wave function develops a node.
This is because of the structure of the wave function derived 
in Eq.~(\ref{eq4}) from where one can realize that again this stems from 
the Pauli blocking factor.
The maximum of the wave function shifts to higher momenta and 
follows the increase of the Fermi momentum $k_F$, as indicated 
on Fig.~\ref{fig3}.
From a certain point on the denominator in (\ref{eq9}) develops a zero 
and no self-consistent solution can be found any longer.

On the other hand, the wave functions in position space 
in Figs.~\ref{fig3}-(b2), (b3) and (b4) develop 
an oscillatory behavior, as the chemical potential increases.
This is reminiscent to what happens in BCS theory for the pair wave 
function in position space \cite{m06}.

\section{Discussion and Conclusions}

In this work we again took up the study of the critical temperature 
of $\alpha$-particle (quartet) condensation 
in homogeneous symmetric nuclear matter.
We essentially confirm the behavior of two previous studies 
\cite{beyer,rss98}. 
The objective of the paper was to show that practically same results as before 
can be obtained with a strongly simplifying ansatz for the four particle wave 
function. Namely, this time, we used a momentum projected mean field 
variational wave function. This is based on the fact that the four 
different fermions of the quartet can occupy 
the same single particle $0S$-wave function in the mean field. The 
latter is to be determined 
from a self-consistent non linear HF-type of equation as a function of 
chemical potential or density. 
The relation between the chemical potential and density is taken from 
the free Fermi gas relation, Eq.~(\ref{eq-density}). 
%The situation is analogous to the one of the Thouless 
%criterion for ordinary pairing. 
%This means that we cannot attain the BEC limit for the 
%critical temperature in the present formulation. 
%However, the total density of this system 
%must be calculated from the sum of free nucleon density represented by 
%Fermi distribution function
%and $\alpha$-particle density (multiplied by $4$) 
%represented by Bose distribution function.
However, the total nucleon density of the system must be calculated 
from the mean single nucleon state occupation number 
taking into account correlations,
so that the contribution of bound states to the total nucleon density 
is taken into account, see Ref.~\cite{srs90}.
To calculate the critical temperature
not as function of the free nucleon density,
see Fig.~\ref{fig2}b, but of the total nucleon density,
a generalization \`a la NSR \cite{nsr85} must be performed,
%For this a generalization \`a la NSR \cite{nsr85} must be performed, 
that is we have at least to incorporate the 
contribution of the $\alpha$-particle density including the
condensate to the single particle occupation numbers. This shall be 
investigated in future work.

Besides, in this work, we used the isospin-independent 
separable potential, Eq.~(\ref{eq5}), for the two-body interaction 
as a simplification.
Comparison with a realistic two-body interaction is interesting. 
This study also shall be done in the future.

The self-consistent 
wave function has been studied in momentum and position space. 
For negative chemical 
potential the single particle wave function behaves like a Gaussian. 
However, once the 
chemical potential turns positive, then the single particle wave function 
in $r$-space starts to oscillate. This is a well known feature from 
ordinary pairing.

We, therefore, have demonstrated that a very simplifying momentum 
projected mean field 
ansatz suffices to account for the salient features of quartet 
condensation. This is very helpful for the next step which is more 
complicated, i.e. the incorporation of quartet condensation 
self-consistently into the Equation of State (EOS).

We should, however, be aware of the fact that our projected mean field ansatz 
for the quartet wave function can only be a valid approximation so long as  
well defined quartets exist. In the break down region seen on 
Fig.~\ref{fig2}, this is 
certainly no longer the case. How the quartet phase evolves into a superfluid 
phase of pairs is an open question. A possibility to study this very 
interesting problem could be to write down the in-medium four body 
equation (\ref{eq1}) directly in the BCS formalism, i.e. with 
the corresponding BCS coherence factors. It may be foreseen that the 
latter only catch on close to 
the transition region. Another interesting problem for the future is how the 
present results are modified in the asymmetric case, that is in the case of 
neutron excess.

The success of our study to employ a very simplifying ansatz of the mean 
field type for the quartet wave function, may open wide perspectives. 
Besides to push the description 
of quartet condensation much further, 
there might exist the possibility 
that even for the case of a gas of trions such a 
projected mean field ansatz is a quite valid approach. In the case of 
three colors, like quarks in the constituent quark model for nucleons, 
a harmonic 
confining potential is frequently assumed and the three quarks can occupy the 
lowest $0S$ state, analogously to the case of quartets treated in the present 
paper. Of course, trions are composite fermions and cannot be 
treated in the same way as 
bosonic composites, since they shall form a new Fermi gas with their own new
Fermi level. How this situation can eventually be treated has recently been 
outlined in \cite{ssr08}.

\acknowledgments

This work is part of an ongoing collaboration with Y. Funaki, H. Horiuchi, 
A. Tohsaki, and T. Yamada. Useful discussions are gratefully acknowledged.
We thank P. Nozi\`eres for his interest in quartet condensation.
This work is supported by the DFG grant No. RO905/29-1.


\begin{thebibliography}{99}

\bibitem{sb08}N. Sandulescu and G. F. Bertsch, arXiv:0808.3009; 
N. Pillet, N. Sandulescu, and P. Schuck, 
Phys. Rev. C {\bf 76}, 024310 (2007).

\bibitem{gps08}For a review, S. Giorgini, L. P. Pitaevskii, and S. Stringari,
Rev. Mod. Phys. {\bf 80}, 1215 (2008).


\bibitem{olk08}T. B. Ottenstein, T. Lompe, M. Kohnen, A. N. Wenz, 
and S. Jochim,
Phys. Rev. Lett. {\bf 101}, 203202 (2008);
J. H. Huckans, J. R. Williams, E. L. Hazlett, R. W. Stites, and K. M. O'Hara,
arXiv:0810.3288.


\bibitem{km05}H. Kamei and K. Miyake, 
J. Phys. Soc. Jpn. {\bf 74}, 1911 (2005).

\bibitem{sg99}A. S. Stepanenko, J. M. F. Gunn, 
arXiv: cond-mat/9901317.


\bibitem{rcl08}
P. Schlottmann, J. Phys. Condens. Matter {\bf 6}, 1359 (1994);
C. Wu, Phys. Rev. Lett. {\bf 95}, 266404 (2005);
G. Roux, S. Capponi, P. Lecheminant, and P. Azaria,
Eur. Phys. J. B (2008).


\bibitem{www37}
E. P. Wigner, Phys. Rev. {\bf 51}, 106 (1937); 
J. A. Wheeler, Phys. Rev. {\bf 52}, 1083 (1937); 
W. Wefelmeier, Z. Phys. {\bf 107}, 332 (1937).


\bibitem{rss98}
G. R\"opke, A. Schnell, P. Schuck, and P. Nozi\`eres,
Phys. Rev. Lett. {\bf 80}, 3177 (1998).


\bibitem{thsr01}A. Tohsaki, H. Horiuchi, P. Schuck, and G. R\"opke, 
Phys. Rev. Lett. {\bf 87}, 192501 (2001);
Y. Funaki, T. Yamada, H. Horiuchi, G. R\"opke, P. Schuck, and A. Tohsaki, 
Phys. Rev. Lett. {\bf 101}, 082502 (2008).


\bibitem{st83}S. L. Shapiro and  S. A. Teukolsky, 
{\it Black holes, white Dwarfs and Neutron Stars: 
The Physics of Compact Objects} 
(Wiley, N.Y., 1983); 
D. Pines, R. Tamagaki, and  S. Tsuruta (eds.), {\it Neutron Stars}
(Addison-Weseley, N.Y., 1992). 

\bibitem{sto98}H. Shen, H. Toki, K. Oyamatsu, and  K. Simyoshi, 
Prog. Theor. Phys. {\bf 100}, 1013 (1998).


\bibitem{beyer} M. Beyer, Few Body Syst. {\bf 31}, 151 (2002).


\bibitem{nsr85}P. Nozi\`eres and S. Schmitt-Rink,
J. Low Temp. Phys. {\bf 59}, 195 (1985).


\bibitem{rs00}P. Ring and P. Schuck, 
{\it The Nuclear Many-Body Problem} 
(Springer-Verlag, Berlin Heidelberg New York, 2000).

\bibitem{berger}J.-F. Berger, private communication.

\bibitem{ds94}
J. Eichler, T. Marumori, and K. Takada, 
Prog. Theor. Phys. {\bf 40}, 60 (1968);
G. R\"opke, L. M\"unchow, and H. Schulz,
Nucl. Phys. {\bf A379}, 536 (1982);
G. R\"opke, M. Schmidt, L. M\"unchow, and H. Schulz,
Nucl. Phys. {\bf A399}, 587 (1983);
P. Danielewicz and P. Schuck
Nucl. Phys. {\bf A567},  78 (1994).


\bibitem{thouless60}
D. J. Thouless, Ann. Phys. {\bf 10}, 553 (1960).


\bibitem{schuck08}P. Schuck, 
{\it State of the Art in Nuclear Cluster Physics}, Strasbourg, May 2008, 
to appear in Int. J. Mod. Phys. E.

\bibitem{m06} M. Matsuo,
Phys. Rev. C {\bf 73}, 044309 (2006).

\bibitem{srs90}
M. Schmidt, G. R\"opke, and H. Schulz,
Ann. Phys. (N.Y.) {\bf 202}, 57 (1990).


\bibitem{ssr08}P. Schuck, T. Sogo, and  G. R\"opke,
arXiv:0812.1709.


\end{thebibliography}
\end{document}